\documentclass[pra,aps,amsmath,amssymb,amsfonts,twocolumn,nofootinbib,longbibliography,superscriptaddress]{revtex4-2}

\usepackage{bm,mathrsfs}
\usepackage{graphicx}
\usepackage{epsfig}
\usepackage[T1]{fontenc}
\usepackage{amsmath,amsfonts,amssymb,bbm}
\usepackage{times}
\usepackage{newtxtext,newtxmath}
\usepackage{verbatim}
\usepackage[sort&compress]{natbib}
\usepackage[colorlinks,breaklinks,linkcolor=blue,anchorcolor=blue,citecolor=blue,urlcolor=blue]{hyperref}

\allowdisplaybreaks[4]

\begin{document}

\title{Engineering the non-Hermitian Su-Schrieffer-Heeger model with skin effects in Rydberg atom arrays}

\author{J. N. Bai}
\affiliation{Center for Quantum Science and School of Physics, Northeast Normal University, Changchun 130024, China}

\author{F. Yang}
\affiliation{School of Physics and Zhejiang Key Laboratory of Micro-nano Quantum Chips and Quantum Control, Zhejiang University, Hangzhou 310027, China}
\affiliation{Niels Bohr International Academy, Niels Bohr Institute, University of Copenhagen, DK-2100 Copenhagen, Denmark}

\author{D. Yan}
\email{yand@hainnu.edu.cn}
\affiliation{College of Physics and Electronic Engineering, Hainan Normal University, Haikou 571158, China}

\author{Weibin Li}
\affiliation{School of Physics and Astronomy and Centre for the Mathematics and Theoretical Physics of Quantum Non-equilibrium Systems, The University of Nottingham, Nottingham NG7 2RD, United Kingdom}

\author{X. Q. Shao}
\email{xqshao@nenu.edu.cn}
\affiliation{Center for Quantum Science and School of Physics, Northeast Normal University, Changchun 130024, China}
\affiliation{Institute of Quantum Science and Technology, Yanbian University, Yanji 133002, China}

\date{\today}

\begin{abstract}
We propose and systematically analyze a practical scheme for implementing a one-dimensional non-Hermitian (NH) Su-Schrieffer-Heeger model using individually addressable Rydberg atom arrays. Our setup consists of an atomic chain with three-atom unit cells, in which a synthetic gauge field is generated by applying multicolor laser fields. By engineering fast dissipative channels for one auxiliary atom in each unit cell, adiabatic elimination effectively gives rise to a NH skin effect. We examine how fluctuations in the experimental parameters influence both the skin effect and the topological invariant in real space and find that both features remain highly robust. This work establishes a versatile, controllable, and programmable open-system quantum simulator with neutral atoms, providing a clear route for exploring rich NH topological phenomena.
\end{abstract}

\maketitle

\section{Introduction}

Rydberg atoms are neutral atoms excited to high-lying electronic states with large principal quantum numbers. Owing to their strong long-range interactions, large polarizabilities, and long lifetimes, they provide a highly controllable platform for quantum simulation and quantum information processing~\cite{Gallagher_1994,QuantumInformationWithRydbergAtoms,saffman2016quantum,shao2024rydberg}. In optical-tweezer arrays, individual atoms can be assembled, addressed, and detected with single-site resolution, while their interactions can be tuned through dipole-dipole or van der Waals (vdW) forces~\cite{PhysRevLett.112.183002,PhysRevLett.113.193002,PhysRevLett.115.093002,labuhn2016tunable,endres2016atom,PhysRevX.8.011032,PhysRevLett.120.123204,PhysRevLett.124.043402,PhysRevLett.124.130604,PhysRevResearch.3.023008,PhysRevLett.128.013603,ebadi2022quantum,PhysRevLett.128.113602,PRXQuantum.3.010344,PhysRevLett.128.120503,PRXQuantum.3.020303,PhysRevResearch.4.L032046,PhysRevLett.129.243202,PhysRevLett.130.220601,srakaew2023subwavelength,PhysRevLett.131.123201,PhysRevLett.131.203003,evered2023high,ma2023high,zhao2023floquet,PhysRevLett.132.076505,PhysRevB.110.075111,PhysRevResearch.6.013293,PhysRevA.110.042603,PhysRevResearch.6.023031,PhysRevLett.132.206503,PhysRevResearch.7.L012009,5qhh-322q}. These capabilities have enabled the realization of programmable quantum spin models~\cite{chen2023continuous,PhysRevLett.130.243001,bornet2023scalable,PhysRevB.106.115122,PRXQuantum.3.020303,PRXQuantum.4.010316,746s-fv7x}, the study of entanglement generation~\cite{omran2019generation,PhysRevLett.123.230501,PhysRevResearch.3.023021,PhysRevLett.127.090602,graham2022multi,schine2022long,o2023entanglement,PhysRevLett.132.113601}, many-body dynamics~\cite{bernien2017probing,wang2020preparation,bluvstein2021controlling,ebadi2021quantum,PhysRevX.12.021028}, and nonequilibrium quantum phenomena~\cite{omran2019generation,PhysRevLett.122.040603,PhysRevResearch.2.022065,PhysRevX.11.021021,PhysRevLett.127.090602,PhysRevLett.128.090606,PhysRevLett.133.216601,PhysRevLett.134.050403,w1cp-l5vq,4my3-vk6c,PhysRevX.15.011035,PRXQuantum.4.040339}. Beyond conventional spin dynamics, Rydberg platforms have also become increasingly important for exploring topological phases of matter~\cite{de2019observation,semeghini2021probing,khazali2022discrete,PhysRevLett.129.195301,PhysRevLett.130.043601,PhysRevLett.130.206501,yue2026average}. Their programmable interactions make it possible to engineer symmetry-protected topological states as well as states with intrinsic topological order~\cite{PhysRevA.107.062407,PhysRevX.13.031008}, which are of direct relevance to robust quantum information processing. At the same time, dissipation engineering in Rydberg systems has attracted growing attention as a route to controlled nonequilibrium dynamics and steady-state preparation~\cite{PhysRevA.102.053118,PhysRevApplied.20.014014,PhysRevResearch.5.043036,w3x9-ll79,gb6x-m1sg,guo2025scalablesteadystateentanglementfloquetengineered}. The coexistence of coherent interactions, programmable geometry, and controllable dissipation therefore makes Rydberg atom arrays a natural setting for studying topological phenomena beyond closed Hermitian systems.

Non-Hermitian (NH) systems, in which gain, loss, dissipation, or nonreciprocal hopping is effectively incorporated into the system dynamics, exhibit spectral and topological properties with no Hermitian counterpart~\cite{PhysRevLett.116.133903,PhysRevLett.121.026808,PhysRevLett.121.086803,PhysRevLett.123.066404,PhysRevLett.123.097701,PhysRevX.9.041015,PhysRevLett.125.226402,RevModPhys.93.015005,PhysRevLett.126.216407,PRXQuantum.4.030315}. Under symmetry constraints such as parity-time ($\mathcal{PT}$) symmetry, NH systems can support entirely real spectra~\cite{PhysRevLett.80.5243,PhysRevA.106.023309}. More generally, however, they host intrinsic NH phenomena, including exceptional degeneracies, half-integer topological charges, exceptional rings, and the non-Hermitian skin effect (NHSE)~\cite{PhysRevLett.118.045701,PhysRevLett.123.066405,PhysRevLett.129.084301,PhysRevResearch.7.023062,PhysRevLett.124.086801,PhysRevX.13.021007,PhysRevLett.129.070401,xbj1-hfyf}. In particular, the NHSE drives an extensive number of eigenstates to accumulate near the system boundaries, thereby invalidating the conventional Bloch bulk-boundary correspondence and motivating generalized descriptions such as non-Bloch band theory and biorthogonal polarization~\cite{PhysRevLett.123.066404,PhysRevResearch.2.043046}.

Experimentally, NH physics has been implemented on various platforms, including classical circuits, photonic systems, and synthetic dimensions~\cite{lustig2019photonic,dutt2020single,buddhiraju2021arbitrary,kanungo2022realizing,PhysRevA.109.032801,PhysRevA.110.023318,PhysRevA.110.L040601,chen2025interaction}. Real-space Rydberg atom arrays offer several advantages for simulating NH physics beyond existing implementations~\cite{PhysRevA.106.023309,lf2c-h1kk,zhang2025observationnonhermitianmanybodyphase}. First, optical tweezers provide scalable real-space geometries, high-fidelity single-site addressability, and flexible control of boundary conditions, which are essential for directly resolving the boundary accumulation associated with the NHSE and for probing the influence of local disorder~\cite{evered2023high,ma2023high}. Second, unlike static single-particle photonic implementations, Rydberg atoms allow dissipation to be actively engineered through laser-driven atomic transitions, providing dynamical tunability and state selectivity that are difficult to realize in macroscopic gain-loss media~\cite{PhysRevLett.122.053601,PhysRevLett.131.080403}. Finally, the combination of software-reconfigurable tweezer geometries and strong long-range Rydberg interactions opens a natural route toward higher-dimensional and interacting many-body NH systems~\cite{keesling2019quantum,scholl2021quantum}.

In this work, we propose a scheme for engineering nonreciprocal Su-Schrieffer-Heeger (SSH) physics in a real-space Rydberg atom array by combining synthetic gauge fields with dissipation-assisted excitation exchange. Specifically, we dress individually addressable atoms with multicolor laser fields~\cite{PhysRevLett.123.063001,PhysRevLett.125.143601,PhysRevResearch.4.L032046,PRXQuantum.1.020303} to control the photon-assisted spin-exchange interactions, while auxiliary atoms are introduced to mediate effective directional hopping through adiabatic elimination. This construction enables independently tunable intracell and intercell nonreciprocal couplings in a scalable optical-tweezer array.

The remainder of this paper is organized as follows. In Sec.~\ref{sec2}, we introduce the basic building block consisting of a three-atom unit cell and analyze the effective dynamics in the single-excitation subspace. We compare the original Hamiltonian with the reduced effective Hamiltonian and clarify the roles of the intracell and intercell excitation-exchange processes. In Sec.~\ref{sec3}, we show how fast dissipative channels of the auxiliary atoms can be used to generate nonreciprocal couplings for both intracell and intercell processes, and construct a scalable nonreciprocal SSH chain composed of $N$ unit cells and study its properties under open boundary conditions (OBCs). In Sec.~\ref{sec5}, we analyze the effects of experimentally relevant noise sources and further examine the robustness of the bulk and edge states against these disorders by evaluating the skin-effect order parameter and topological invariant. In Sec.~\ref{sec6}, we extend the analysis to the corresponding nonreciprocal SSH model under periodic boundary conditions (PBCs). Finally, Sec.~\ref{sec7} summarizes the main results and discusses the outlook of this work.

\section{Three-atom unit-cell chain model}\label{sec2}

\begin{figure}
\includegraphics[width=1\linewidth]{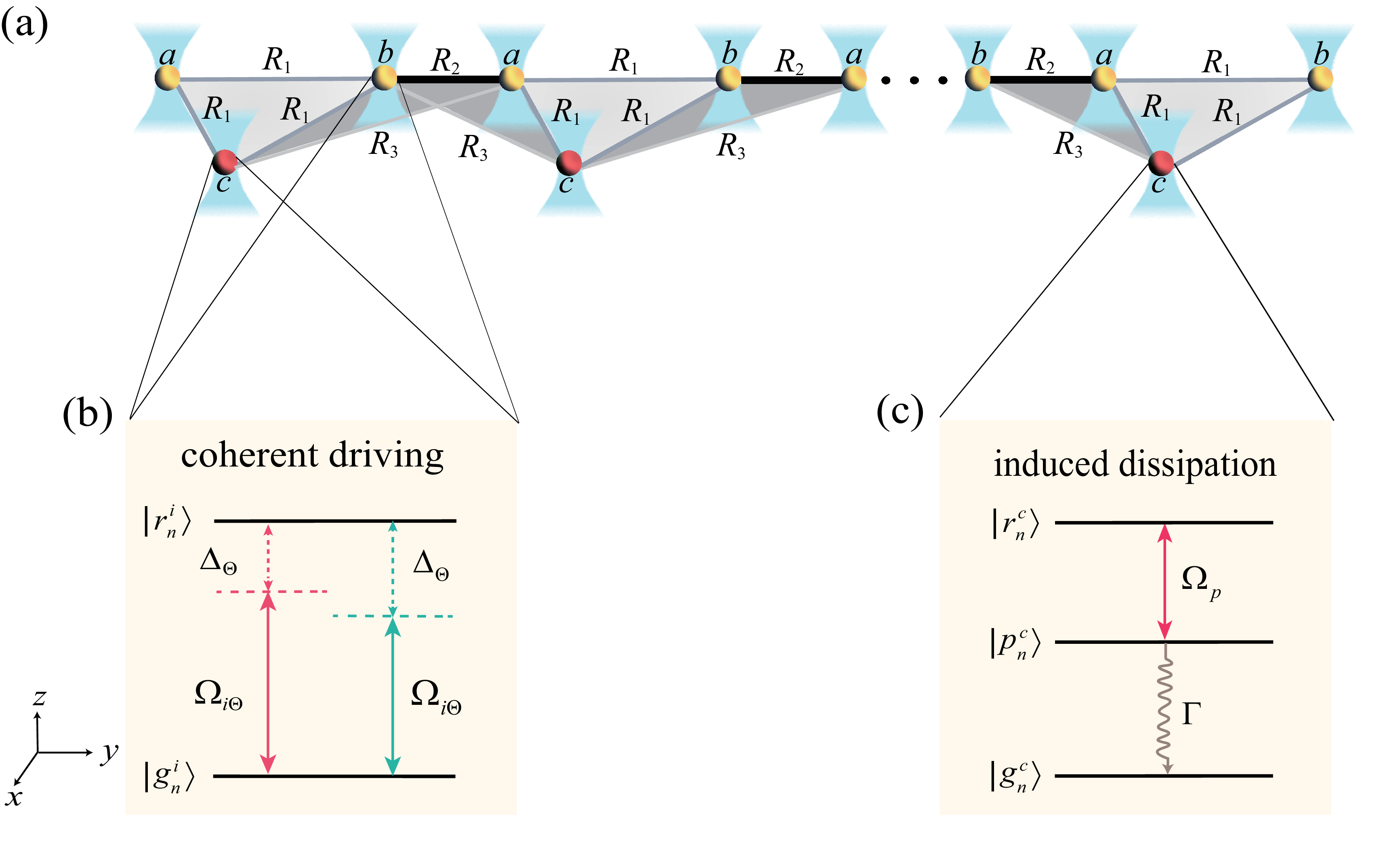}
\caption{(a) The schematic of a chain structure consisting of $N$ three-atom unit cells. Each unit cell contains two data atoms (yellow) and one auxiliary atom (red), with the distances between atoms inside the unit cell labeled as $R_1=6~\mu\mathrm{m}$ and distances between the adjacent unit cells denoted by $R_2=3.46~\mu\mathrm{m}$ and $R_3=8.29~\mu\mathrm{m}$. (b) Each atom is driven by a two-color laser field~\cite{PhysRevResearch.4.L032046}, and each nearest-neighbor pair within the triangular plaquette shares one frequency component to address the corresponding intracell coupling channel. (c) A fast dissipative channel is constructed by introducing an intermediate state of the auxiliary atoms.}\label{fig1}
\end{figure}

We consider a Rydberg-atom array composed of $N$ unit cells. Each unit cell hosts three individually addressable $^{87}\mathrm{Rb}$ atoms arranged at the vertices of an equilateral triangle with a side length $R_1$, as illustrated in Fig.~\ref{fig1}(a). The $j$th atom in the $n$th cell is driven by two laser fields (a two-color drive) characterized by Rabi frequencies $\Omega^j_{\Theta}=|\Omega^j_{\Theta}|e^{i\varphi^j_{\Theta}}$ and detunings $\Delta^j_\Theta$ [see Fig.~\ref{fig1}(b)], which couple the ground state $\lvert g_n^j \rangle=|5S_{1/2},F=2,m_F=2\rangle$ to the Rydberg state $\lvert r_n^j \rangle=|70S_{1/2},m_J=1/2\rangle$. Here, the index $\Theta$ labels the available lasers, $\Theta\in\{\mathrm{I},\mathrm{II},\mathrm{III}\}$. The phase $\varphi^j_{\Theta}$ specifies the optical phase of the laser addressing the $j$th atom. In the interaction picture, the system Hamiltonian takes the form
\begin{eqnarray}\label{eq1}
H_I &=& \sum_{n=1}^{N}\sum_{j=a,b,c}\sum_{\Theta}
\left[
\frac{\Omega^j_{\Theta}}{2}e^{i\Delta^j_{\Theta}t}
\lvert r_n^j\rangle\langle g_n^j\rvert+\mathrm{H.c.}
\right]
\nonumber\\
&&+\sum_{\{(n,j),(m,k)\}}V_{n,m}^{jk}
\lvert r_n^j r_m^k\rangle\langle r_n^j r_m^k\rvert.
\end{eqnarray}
Here, the summation $\sum_{\{(n,j),(m,k)\}}$ runs over all unordered pairs of distinct atoms retained in the model, so that each vdW interaction is counted only once. The interaction strength $V_{n,m}^{jk}=-C_6/(R_{n,m}^{jk})^6$ denotes the vdW interaction between atom $j$ in unit cell $n$ and atom $k$ in unit cell $m$, and $R_{n,m}^{jk}$ is the corresponding distance. $C_6/(2\pi)\approx-863~\mathrm{GHz}~(\mu\mathrm{m})^6$ is the vdW dispersion coefficient of the Rydberg state~\cite{vsibalic2017arc}. In each unit cell, $a$ and $b$ label the data atoms, while $c$ denotes the auxiliary atom.

In the large-detuning regime $|\Omega_{\Theta}^{j}|\ll|\Delta_{\Theta}^{j}|$, an effective hopping amplitude $J_{n,m}^{jk}$ is obtained to describe the laser-assisted dipolar exchange between atom $j$ in unit cell $n$ and atom $k$ in unit cell $m$, i.e., $|r_n^j g_m^k\rangle\leftrightarrow|g_n^j r_m^k\rangle$. It reads~\cite{PhysRevLett.123.063001,PhysRevResearch.4.L032046}
\begin{equation*}
J_{n,m}^{jk}
=
\left|
\frac{\Omega_{\Theta^{[jk]}}^{j}\Omega_{\Theta^{[jk]}}^{k*}V_{n,m}^{jk}}
{4\Delta_{\Theta^{[jk]}}\left(\Delta_{\Theta^{[jk]}}+V_{n,m}^{jk}\right)}
\right|e^{i\phi^{jk}_{\Theta^{[jk]}}}.
\end{equation*}
For $m=n$, this expression reduces to the intracell hopping amplitude $J_{n,n}^{jk}$. To suppress the off-resonant crosstalk, the energy mismatch between different colored lasers must strictly satisfy $\delta\Delta=|\Delta_{\Theta}^j-\Delta_{\Theta'}^k|\gg J_{n,m}^{jk}$~\cite{PhysRevResearch.4.L032046}. Owing to the independent phase control of all dressing laser fields, a synthetic magnetic flux can be precisely threaded through a triangular plaquette in each unit cell. The effective gauge flux enclosed by this loop is $\Phi=(\varphi^a_{\rm I}-\varphi^b_{\rm I})+(\varphi^b_{\rm II}-\varphi^c_{\rm II})+(\varphi^c_{\rm III}-\varphi^a_{\rm III})$, where $\varphi^j_{\Theta^{[jk]}}$ and $\varphi^k_{\Theta^{[jk]}}$ are the phases imprinted by the same laser beam $\Theta^{[jk]}$ that mediates the dipolar interaction between sites $j$ and $k$ (with $\Theta^{[ab]}=\mathrm{I}$, $\Theta^{[bc]}=\mathrm{II}$, and $\Theta^{[ca]}=\mathrm{III}$). The resulting phase difference $\phi^{jk}_{\Theta^{[jk]}}=\varphi^j_{\Theta^{[jk]}}-\varphi^k_{\Theta^{[jk]}}$ enters the effective hopping amplitude as a Peierls phase factor. For simplicity, by setting $\varphi^c_{\rm III}=\pm\pi/2$ and all other phases of the laser field to zero, a net flux $\Phi=\pm\pi/2$ is generated through the triangular plaquette, resulting in chiral excitation transport~\cite{li2022coherent}.

\begin{figure}
\includegraphics[width=1\linewidth]{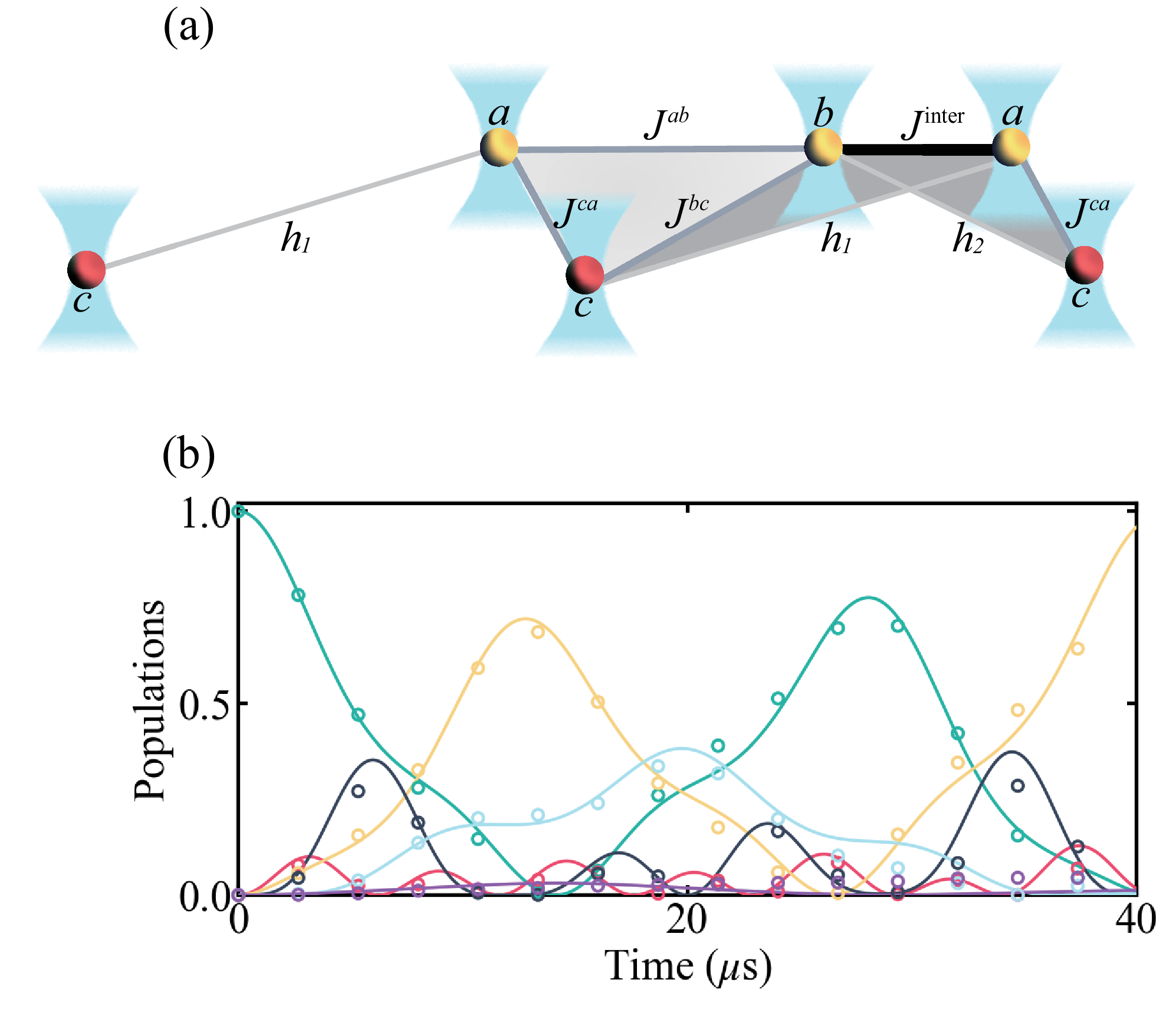}
\caption{(a) Illustration of the six-atom model. (b) Time evolution of the Rydberg-state populations $P_{n-1,c}=\langle\sigma_{n-1,c}^+\sigma_{n-1,c}^-\rangle$ (purple), $P_{n,a}=\langle\sigma_{n,a}^+\sigma_{n,a}^-\rangle$ (green), $P_{n,b}=\langle\sigma_{n,b}^+\sigma_{n,b}^-\rangle$ (red), $P_{n,c}=\langle\sigma_{n,c}^+\sigma_{n,c}^-\rangle$ (yellow), $P_{n+1,a}=\langle\sigma_{n+1,a}^+\sigma_{n+1,a}^-\rangle$ (black), and $P_{n+1,c}=\langle\sigma_{n+1,c}^+\sigma_{n+1,c}^-\rangle$ (blue) governed by Eqs.~(\ref{eq1}) (open circles) and (\ref{eq2}) (solid lines) from the initial state $\sigma_{n,a}^+|G_6^{\rm full}\rangle$. The detunings are $(\Delta_{\rm I},\Delta_{\rm II},\Delta_{\rm III})=2\pi\times(51.3,59.8,68.4)~\mathrm{MHz}$, with Rabi frequencies $(\Omega_{\rm I},\Omega_{\rm II},\Omega_{\rm III})=2\pi\times(4.3,4.65,5)~\mathrm{MHz}$.}\label{fig2}
\end{figure}

Since the Rydberg interaction is distance dependent, couplings among all atoms should, in principle, be taken into account. However, because the vdW interaction decreases rapidly with increasing separation, only a finite number of dominant interaction channels needs to be retained in practice. To justify this approximation, we first consider a six-atom model representing a local segment of the array, as shown in Fig.~\ref{fig2}(a). This segment extends from the $c$ atom in the $(n-1)$th unit cell to the $a$ and $c$ atoms in the $(n+1)$th unit cell, with $n\in[2,N-1]$, and therefore contains the relevant atoms from three consecutive unit cells. This local model captures all coupling processes entering the effective description, including intracell couplings, nearest-neighbor intercell couplings, and residual couplings between auxiliary atoms and neighboring data atoms. The corresponding all-ground-state configuration is denoted by
$
|G_6^{\rm full}\rangle
=|g_{n-1}^{c},g_n^{a},g_n^{b},g_n^{c},g_{n+1}^{a},g_{n+1}^{c}\rangle,
$
which serves as the reference ground state of the considered six-atom model.

In the following effective Hamiltonian, we use simplified symbols for the dominant couplings: $J^{ab}\equiv J_{n,n}^{ab}$, $J^{bc}\equiv J_{n,n}^{bc}$, and $J^{ca}\equiv J_{n,n}^{ca}$ for the intracell couplings; $J^{\rm inter}\equiv J_{n,n+1}^{ba}$ for the designed intercell coupling; and $h_1\equiv J_{n-1,n}^{ca}=J_{n,n+1}^{ca}$ and $h_2\equiv J_{n,n+1}^{bc}$ for the residual nearest-neighbor couplings involving auxiliary atoms. In deriving the effective Hamiltonian below, we take the specific gauge choice $\phi_{\rm III}^{ca}=\pi/2$ for the laser-assisted $c\leftrightarrow a$ coupling. Under this choice, the corresponding $c\leftrightarrow a$ hopping terms acquire a phase factor $e^{i\phi_{\rm III}^{ca}}=i$. In the single-excitation manifold, the resulting effective Hamiltonian can then be written as
\begin{equation}\label{eq2}
H_{\rm eff}=H_{\rm hop}+H_{\rm Stark},
\end{equation}
with effective hopping interaction
\begin{eqnarray}\label{hop}
H_{\rm hop}&=&J^{ab}\sigma_{n,a}^+\sigma_{n,b}^-+J^{bc}\sigma_{n,b}^+\sigma_{n,c}^-\nonumber\\
&&+iJ^{ca}\left(\sigma_{n,c}^+\sigma_{n,a}^-+\sigma_{n+1,c}^+\sigma_{n+1,a}^-\right)\nonumber\\
&&+J^{\rm inter}\sigma_{n,b}^+\sigma_{n+1,a}^-+h_2\sigma_{n,b}^+\sigma_{n+1,c}^-\nonumber\\
&&+ih_1\left(\sigma_{n-1,c}^+\sigma_{n,a}^-+\sigma_{n,c}^+\sigma_{n+1,a}^-\right)+\mathrm{H.c.},
\end{eqnarray}
and the Stark shifts
\begin{eqnarray}\label{stark}
H_{\rm Stark}&=&\mu^c_{n-1}\sigma_{n-1,c}^+\sigma_{n-1,c}^-+\sum_{j=a,b,c}\mu^j_n\sigma_{n,j}^+\sigma_{n,j}^-\nonumber\\
&&+\sum_{j=a,c}\mu^j_{n+1}\sigma_{n+1,j}^+\sigma_{n+1,j}^-.
\end{eqnarray}
Here, $\mu_n^j$ denotes the effective Stark shift of atom $j$ in the $n$th unit cell within the considered finite-range model, which includes the single-particle ac Stark shift and the interaction-dependent Stark-shift corrections induced by atoms in both the same and different unit cells. Explicitly,
$\mu_n^j=\sum_{\Theta}|\Omega_{\Theta}^{j}|^2/(4\Delta_{\Theta}^{j})
-\sum_{k\neq j}\sum_{\Theta}|\Omega_{\Theta}^{k}|^2/[4(\Delta_{\Theta}^{k}+V_{n,n}^{jk})]
-\sum_{m\neq n}\sum_k\sum_{\Theta}|\Omega_{\Theta}^{k}|^2/[4(\Delta_{\Theta}^{k}+V_{n,m}^{jk})]$.
The operators $\sigma_{n,j}^{+}=|r_n^j\rangle\langle g_n^j|$ and $\sigma_{n,j}^{-}=|g_n^j\rangle\langle r_n^j|$ denote the local raising and lowering operators for atom $j$ in the $n$th unit cell; their action on all other atoms is implicitly taken to be the identity. In Fig.~\ref{fig2}(b), we compare the site-resolved population dynamics obtained from Eqs.~\eqref{eq1} and \eqref{eq2}. The two descriptions show excellent quantitative agreement, confirming that interactions beyond the cutoff distance $R_3$ can be safely neglected in subsequent simulations. In addition, the Stark-shift terms appearing in Eq.~\eqref{stark} can be compensated by suitable parameter tuning or absorbed into the effective detunings. We therefore omit these terms in the following discussion.

\section{Nonreciprocal coupling between atoms}\label{sec3}

\subsection{Induced dissipation of the Rydberg state}

Given the weak spontaneous emission rate $\gamma^c$ of the Rydberg state, constructing an NH channel by directly adiabatically eliminating the auxiliary atom is unfeasible. To address this, we engineer a rapid relaxation process that transfers the auxiliary atom in each unit cell from the Rydberg state $|r^c\rangle$ to the target ground state $|g^c\rangle$ via a short-lived intermediate state $|p^c\rangle$, as shown in Fig.~\ref{fig1}(c). The dynamics of the subsystem are governed by two competing processes: coherent Rabi oscillations between the Rydberg and intermediate states driven by a Rabi frequency $\Omega_p$ and incoherent spontaneous emission from the intermediate state to the ground state at a rate $\Gamma$. The optimization of the relaxation rate requires balancing these processes.

The Liouvillian superoperator $\mathcal{L}$ of this three-level model can be expressed as
\begin{eqnarray}\label{eq3}
\mathcal{L}&=&-i\left(H_{\rm rp}\otimes\mathcal{I}-\mathcal{I}\otimes H_{\rm rp}^{\rm T}\right)+L_{\rm gp}\otimes L_{\rm gp}^{*}\nonumber\\
&&-\frac{1}{2}\left(L_{\rm gp}^{\dagger}L_{\rm gp}\otimes\mathcal{I}+\mathcal{I}\otimes L_{\rm gp}^{\rm T}L_{\rm gp}^{*}\right),
\end{eqnarray}
where $H_{\rm rp}=\Omega_p|p^c\rangle\langle r^c|/2+\mathrm{H.c.}$, $\mathcal{I}$ is the identity operator, and $L_{\rm gp}=\sqrt{\Gamma}|g^c\rangle\langle p^c|$. The superscripts $\rm T$ and $*$ denote the transpose and complex conjugation, respectively. The time evolution of the system is given by
\begin{equation}\label{eq4}
\rho(t)=e^{\mathcal{L}t}\rho_{\rm in}=\rho_{\rm ss}+\sum_{i=1}^{8}a_i e^{\lambda_i t}\mathbb{R}_i,
\end{equation}
where $\rho_{\rm ss}$ is the unique steady state, $\{\lambda_i\}$ are the complex eigenvalues of $\mathcal{L}$, and $a_i=\mathrm{Tr}[\mathbb{L}_i\rho_{\rm in}]$ represents the projection of the initial state $\rho_{\rm in}$ onto the $i$th right eigenmatrix $\mathbb{R}_i$ (associated with the left eigenmatrix $\mathbb{L}_i$). The relaxation timescale of the system is determined by the Liouvillian gap $g$, which is defined as the magnitude of the nonzero eigenvalue with the smallest absolute real part~\cite{PhysRevResearch.5.043036,guo2025scalablesteadystateentanglementfloquetengineered}. Using the basis states $|g^c\rangle\equiv(1~0~0)^{\rm T}$, $|p^c\rangle\equiv(0~1~0)^{\rm T}$, and $|r^c\rangle\equiv(0~0~1)^{\rm T}$ and assuming the initial state lies within the subspace $\{|r^c\rangle,|p^c\rangle\}$, we derive the spectral gap as
\begin{equation}\label{eq5}
g=\mathrm{Re}\left[\frac{1}{2}(\Gamma-\kappa)\right],
\end{equation}
where $\kappa=\sqrt{\Gamma^2-4\Omega_p^2}$.

The behavior of the spectral gap depends critically on the ratio of $\Omega_p$ to $\Gamma$. In the weak-driving regime ($\Omega_p<\Gamma/2$), $\kappa$ is real and positive, yielding a gap of $g=(\Gamma-\kappa)/2$. Standard Rydberg experiments typically operate in this regime to facilitate the adiabatic elimination of the intermediate state, resulting in an effective relaxation rate of $\Omega_p^2/\Gamma$~\cite{w3x9-ll79,gb6x-m1sg}. Although significantly faster than the natural decay of Rydberg states, this rate is inherently limited by the validity condition of the adiabatic approximation ($\Omega_p\ll\Gamma$) and is insufficient for the auxiliary atom elimination required here. To overcome this limitation, we explore the strong-driving regime ($\Omega_p\geq\Gamma/2$). In this limit, $\kappa$ becomes purely imaginary, resulting in a gap of $g=\Gamma/2$. Physically, the strong Rabi drive saturates the effective spontaneous emission of the Rydberg state at half the decay rate of the intermediate $p$ state. This enhancement allows the spectral gap of the NH band structure to reach its global maximum at the Liouvillian exceptional point~\cite{guo2025scalablesteadystateentanglementfloquetengineered}.

\subsection{Adiabatic elimination creates nonreciprocal coupling}

Taking into account the spontaneous decay of the data atoms at rate $\gamma^a=\gamma^b=\gamma$, the system dynamics is governed by a Markovian master equation,
\begin{eqnarray}\label{eq6}
\dot{\rho}&=&-i\left(H'_{\rm eff}\rho-\rho H_{\rm eff}^{\prime\dagger}\right)
+\sum_{\alpha\in\{n,n+1\}}L_{\rm gr}^{(\alpha,a)}\rho L_{\rm gr}^{(\alpha,a)\dagger}\nonumber\\
&&+L_{\rm gr}^{(n,b)}\rho L_{\rm gr}^{(n,b)\dagger}
+\sum_{\alpha\in\{n-1,n,n+1\}}L_{\rm gr}^{(\alpha,c)}\rho L_{\rm gr}^{(\alpha,c)\dagger},
\end{eqnarray}
where the NH effective Hamiltonian is defined as
\begin{eqnarray}\label{eq7}
H'_{\rm eff}&=&H_{\rm hop}-\frac{i}{2}\Bigg(
\sum_{\alpha\in\{n,n+1\}}L_{\rm gr}^{(\alpha,a)\dagger}L_{\rm gr}^{(\alpha,a)}\nonumber\\
&&
+L_{\rm gr}^{(n,b)\dagger}L_{\rm gr}^{(n,b)}+\sum_{\alpha\in\{n-1,n,n+1\}}L_{\rm gr}^{(\alpha,c)\dagger}L_{\rm gr}^{(\alpha,c)}\Bigg),
\end{eqnarray}
with jump operators $L_{\rm gr}^{(\alpha,a)}=\sqrt{\gamma}|g^a\rangle\langle r^a|$, $L_{\rm gr}^{(n,b)}=\sqrt{\gamma}|g^b\rangle\langle r^b|$, and $L_{\rm gr}^{(\alpha,c)}=\sqrt{\Gamma/2}|g^c\rangle\langle r^c|$. Here, we have neglected the spontaneous decay rate $\gamma^c$ of the auxiliary atom since $\gamma^c\ll\Gamma/2$.

\begin{figure}
\includegraphics[width=1\linewidth]{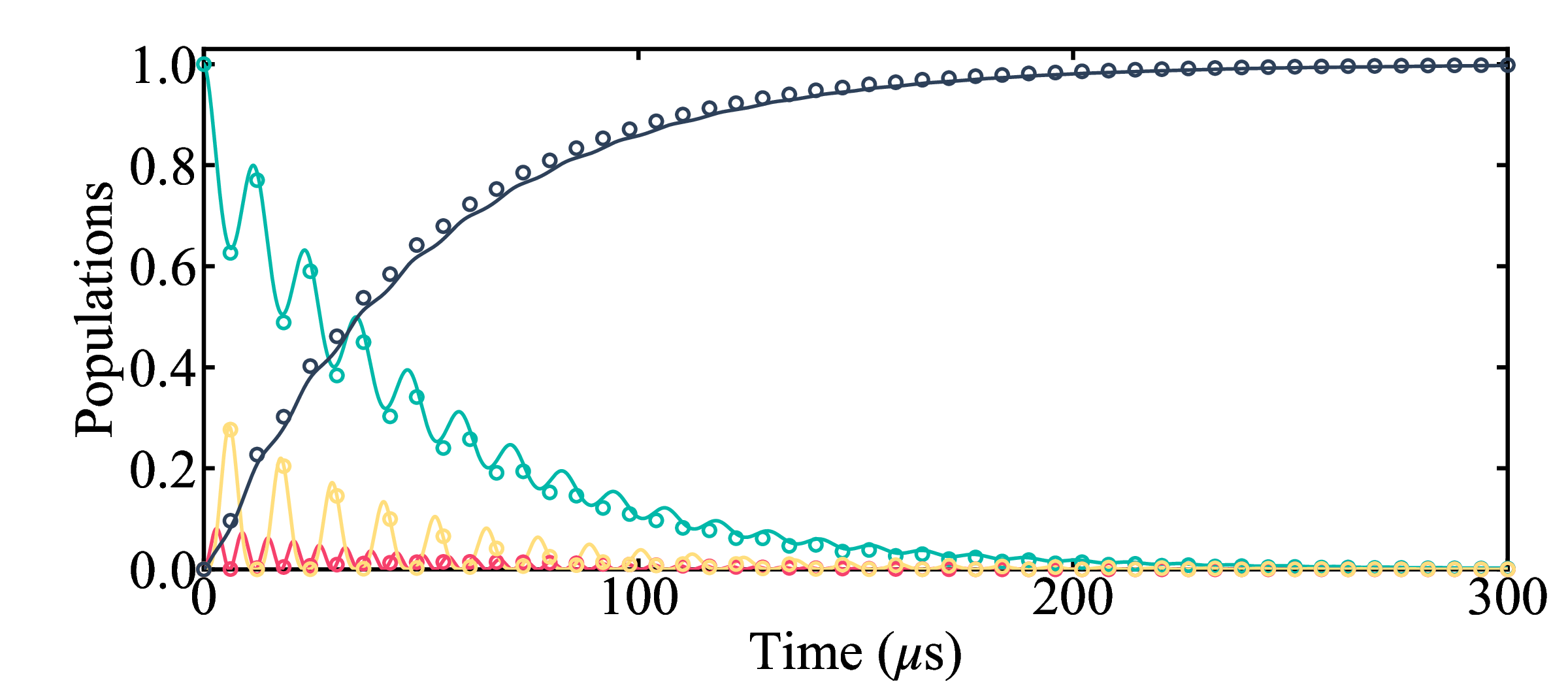}
\caption{Population dynamics initialized with a single excitation on atom $a$ of the $n$th unit cell. Solid lines and open circles denote the results of Eqs.~(\ref{eq6}) and (\ref{10}), respectively. The green, red, and yellow data correspond to the Rydberg-state populations $P_{n,a}$, $P_{n,b}$, and $P_{n+1,a}$, respectively, while the black data denote the ground-state population. The decay rates are $\Gamma/(2\pi)\approx1.35~\mathrm{MHz}$ and $\gamma/(2\pi)\approx1.02~\mathrm{kHz}$.}\label{fig3}
\end{figure}

Within the Hamiltonian framework of Eq.~(\ref{eq7}), we restrict the dynamics to the single-excitation subspace. The corresponding state can then be expanded as $|\Psi(t)\rangle=\sum_n\sum_{j\in\{a,b,c\}}u_n^j(t)\sigma_{n,j}^+|G_6^{\rm full}\rangle$, where $u_n^j(t)$ is the probability amplitude for finding the excitation on the $j$th atom in the $n$th unit cell. Substituting this ansatz into the effective Schr\"odinger equation $i\partial_t|\Psi(t)\rangle=H'_{\rm eff}|\Psi(t)\rangle$, we obtain the following coupled equations for the amplitudes $u_n^j(t)$:
\begin{subequations}\label{8}
\begin{equation}
\dot u^c_{n-1}=-\frac{\Gamma}{4}u_{n-1}^c+h_1u_n^a,
\end{equation}
\begin{equation}
\dot u^a_n=-\frac{\gamma}{2}u_n^a-h_1u_{n-1}^c-iJ^{ab}u_n^b-J^{ca}u_n^c,
\end{equation}
\begin{equation}
\dot u^b_n=-\frac{\gamma}{2}u_n^b-iJ^{ab}u_n^a-iJ^{bc}u_n^c-iJ^{\rm inter}u_{n+1}^a-ih_2u_{n+1}^c,
\end{equation}
\begin{equation}
\dot u^c_n=-\frac{\Gamma}{4}u_n^c+J^{ca}u_n^a-iJ^{bc}u_n^b+h_1u_{n+1}^a,
\end{equation}
\begin{equation}
\dot u^a_{n+1}=-\frac{\gamma}{2}u_{n+1}^a-iJ^{\rm inter}u_n^b-h_1u_n^c-J^{ca}u_{n+1}^c,
\end{equation}
\begin{equation}
\dot u^c_{n+1}=-\frac{\Gamma}{4}u_{n+1}^c-ih_2u_n^b+J^{ca}u_{n+1}^a.
\end{equation}
\end{subequations}
In the regime $\Gamma/4\gg\{h_1,J^{ab},J^{ca},J^{bc},h_2\}$, the rapidly decaying auxiliary atoms can be eliminated adiabatically. Through this mechanism, we enforce $\dot u^c_{n-1}=\dot u^c_n=\dot u^c_{n+1}\approx0$ and obtain
\begin{subequations}\label{9}
\begin{equation}
u^c_{n-1}=\frac{4}{\Gamma}h_1u_n^a,
\end{equation}
\begin{equation}
u^c_n=\frac{4}{\Gamma}\left(J^{ca}u_n^a-iJ^{bc}u_n^b+h_1u_{n+1}^a\right),
\end{equation}
\begin{equation}
u^c_{n+1}=\frac{4}{\Gamma}\left(-ih_2u_n^b+J^{ca}u_{n+1}^a\right).
\end{equation}
\end{subequations}
Substituting Eq.~\eqref{9} into Eqs.~\eqref{8}(b), \eqref{8}(c), and \eqref{8}(e) yields the resulting effective dynamics:
\begin{subequations}\label{10}
\begin{equation}
\dot u^a_n=-\left[\frac{\gamma}{2}+\frac{4\big(J^{{ca}^2}+h_1^2\big)}{\Gamma}\right]u_n^a-i(J^{ab}-J_1)u_n^b,
\end{equation}
\begin{eqnarray}
\dot u^b_n&=&-\left[\frac{\gamma}{2}+\frac{4\big((J^{{bc}^2}+h_2^2\big)}{\Gamma}\right]u_n^b-i(J^{ab}+J_1)u_n^a\nonumber\\
&&-i(J^{\rm inter}+J_2)u_{n+1}^a,
\end{eqnarray}
\begin{equation}
\dot u^a_{n+1}=-\left[\frac{\gamma}{2}+\frac{4\big((J^{{ca}^2}+h_1^2\big)}{\Gamma}\right]u_{n+1}^a-i(J^{\rm inter}-J_2)u_n^b,
\end{equation}
\end{subequations}
where $J_1\equiv4J^{bc}J^{ca}/\Gamma$ and $J_2\equiv4(J^{bc}h_1+J^{ca}h_2)/\Gamma$.

\begin{figure*}
\includegraphics[width=1\linewidth]{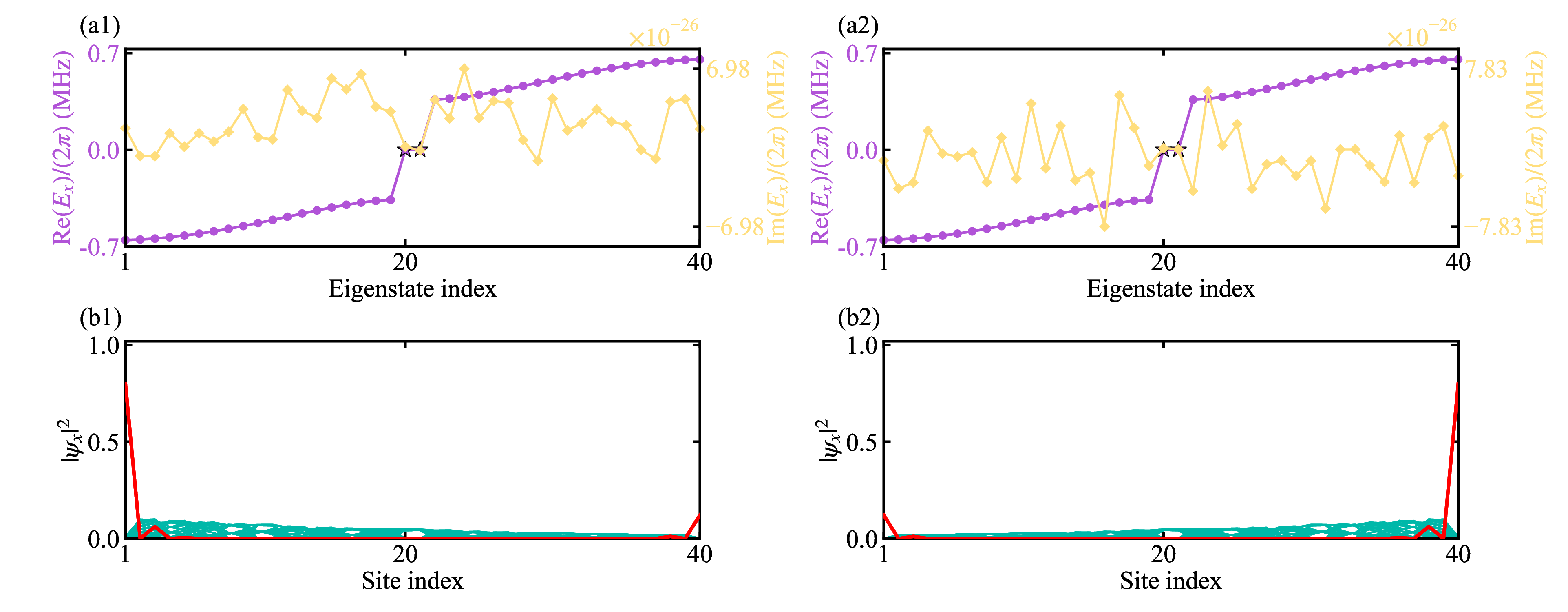}
\caption{(a1) and (a2) show the complex energy spectra of Eq.~(\ref{eq12}) for $\phi^{ca}_{\rm III}=\pi/2$ and $-\pi/2$, respectively. The eigenstates are sorted by increasing $\mathrm{Re}(E_x)$ along the horizontal axis. Purple circles, associated with the left axis, denote $\mathrm{Re}(E_x)/(2\pi)$ (in MHz), while yellow diamonds, associated with the right axis, denote $\mathrm{Im}(E_x)/(2\pi)$ (in MHz). The stars identify the in-gap edge-state eigenvalues near the band-gap center. (b1) and (b2) show the corresponding spatial probability distributions $|\psi_x|^2$, where green curves represent representative extended bulk states and red curves denote boundary-localized edge states. The system size is $L=40$, and the other parameters are the same as in Fig.~\ref{fig3}.}\label{fig4}
\end{figure*}

To verify the validity and effectiveness of dissipation-induced adiabatic elimination of the auxiliary atoms, we compare the dynamical evolutions obtained from Eqs.~(\ref{eq6}) and (\ref{10}). Here, $\Gamma/(2\pi)\approx1.35~\mathrm{MHz}$ and $\gamma/(2\pi)\approx1.02~\mathrm{kHz}$ represent the spontaneous emission rates of the short-lived intermediate state $|p^c\rangle=|6P_{3/2},F=3,m_F=3\rangle$ and the Rydberg state, respectively~\cite{PhysRevApplied.18.044042}. As shown in Fig.~\ref{fig3}, the original and effective descriptions show excellent agreement when we set $\phi^{ca}_{\rm III}=\pi/2$. In addition, with $(J^{{ca}^2}+h_1^2)\approx(J^{{bc}^2}+h_2^2)$, the diagonal dissipative terms in Eq.~(\ref{10}) act as a global isotropic decay proportional to the identity matrix. Although this term sets the overall lifetime of the system, it does not affect the relative amplitudes or the topological structure of the eigenstates and can therefore be gauged away by postselecting the experimental realizations where the excitation survives~\cite{PhysRevA.106.023309}. Finally, we obtain the following NH Hamiltonian, which includes both intracell and intercell couplings:
\begin{eqnarray}\label{eq11}
H_{\rm NH}&=&J_L\sigma^+_{2n-1}\sigma^-_{2n}+J_R\sigma^+_{2n}\sigma^-_{2n-1}\nonumber\\
&&+G_L\sigma^+_{2n}\sigma^-_{2n+1}+G_R\sigma^+_{2n+1}\sigma^-_{2n},
\end{eqnarray}
where $J_L=J^{ab}-J_1$, $J_R=J^{ab}+J_1$, $G_L=J^{\rm inter}+J_2$, and $G_R=J^{\rm inter}-J_2$. Unlike standard $\mathcal{PT}$-symmetric models governed by balanced on-site gain and loss, Eq.~(\ref{eq11}) describes an NH model with nonreciprocal hopping both within ($J_L\neq J_R$) and between ($G_L\neq G_R$) unit cells.

\subsection{NH SSH model under OBCs}\label{sec4}

Under OBCs, we extend the six-atom system to a Rydberg atom array comprising $N=20$ unit cells, thus obtaining the Hamiltonian of the NH SSH model for $L=40$ atoms,
\begin{eqnarray}\label{eq12}
H_{\rm OBC}&=&\sum_{n=1}^{L/2}\left(J_L\sigma^+_{2n-1}\sigma^-_{2n}+J_R\sigma^+_{2n}\sigma^-_{2n-1}\right)\nonumber\\
&&+\sum_{n=1}^{L/2-1}\left(G_L\sigma^+_{2n}\sigma^-_{2n+1}+G_R\sigma^+_{2n+1}\sigma^-_{2n}\right),
\end{eqnarray}
and we restrict our subsequent discussion to the nontrivial topological regime where $G_LG_R>J_LJ_R$~\cite{PhysRevA.95.053626,PhysRevLett.121.086803,PhysRevLett.123.066404}. The direction of the nonreciprocal pumping is controlled by the sign of the Peierls phase. With the gauge choice $\phi^{ca}_{\rm III}=\pi/2$, one obtains the hopping configuration in Eq.~\eqref{eq12}. Choosing the opposite phase $\phi^{ca}_{\rm III}=-\pi/2$ reverses the complex hopping phase and hence reverses the effective nonreciprocity, which is equivalent, under the same gauge convention, to interchanging $J_L\leftrightarrow J_R$ and $G_L\leftrightarrow G_R$.

The spectral properties can be obtained by solving the eigenvalue equation $H_{\rm OBC}|\psi_x\rangle=E_x|\psi_x\rangle$, where $|\psi_x\rangle$ denotes the $x$th eigenstate. Due to the NH nature of the Hamiltonian, the eigenvalues $E_x$ are generally complex. The real part $\mathrm{Re}(E_x)$ corresponds to the effective energy (or oscillation frequency) of the mode, while the imaginary part $\mathrm{Im}(E_x)$ represents the effective decay (or growth) rate induced by the dissipative channels. In Figs.~\ref{fig4}(a1) and \ref{fig4}(a2), we plot the real and imaginary parts of $E_x$, with the state index $x$ sorted in ascending order of $\mathrm{Re}(E_x)$. In this arranged spectrum, the two midgap modes ($x=20$ and 21) correspond precisely to the topological edge states. The distribution of $\mathrm{Re}(E_x)$ shows a clear band gap, suggesting that the bulk behaves as an insulator in terms of the real-energy spectrum. Meanwhile, since the imaginary parts of the eigenvalues satisfy $\mathrm{Im}(E_x)\approx0$, this value indicates that the system remains in the $\mathcal{PT}$-symmetric phase. Interestingly, numerical results indicate that even in this $\mathcal{PT}$-unbroken phase, the eigenstates exhibit exponential localization at the boundaries, revealing the NHSE [Figs.~\ref{fig4}(b1) and \ref{fig4}(b2)].

To understand the topological properties and the bulk-boundary correspondence of the nonreciprocal SSH model under OBCs, we employ the similarity transformation $\widetilde{H}_{\rm OBC}=SH_{\rm OBC}S^{-1}=\widetilde{H}_{\rm OBC}^{\dagger}$ to map the NH system into an equivalent Hermitian form. In this symmetric Hermitian frame, the eigenstates $|\widetilde\psi\rangle$ naturally correspond to extended Bloch waves with uniform amplitude in the lattice, i.e., $|\widetilde\psi(x')|\sim1$. To construct such a transformation, the matrix $S$ is chosen to be diagonal. We set the first element as $s_1=1$, while the last two diagonal elements are given by $s_{2N-1}=(l_1l_2)^{N-1}$ and $s_{2N}=(l_1l_2)^{N-1}l_1$, with the scaling factors $l_1=\sqrt{J_L/J_R}$ and $l_2=\sqrt{G_L/G_R}$. Consequently, the matrix $S$ takes the explicit form
\begin{equation}\label{eq13}
S=\mathrm{diag}\left[1,l_1,l_1l_2,l_1^2l_2,\ldots,(l_1l_2)^{N-1},(l_1l_2)^{N-1}l_1\right].
\end{equation}
To retrieve the physical eigenstates $|\psi_x\rangle$ of $H_{\rm OBC}$, we apply the inverse similarity transformation. As a result, the probability amplitude for our NH SSH model at the $y$th effective lattice site is modulated by the inverse of $S$, yielding
\begin{equation}\label{eq14}
\psi_x(y) \propto s_y^{-1}\tilde{\psi}_x(y)\sim(l_1l_2)^{-y/2}e^{ik'y},
\end{equation}
where $k'$ represents the conventional Bloch wave vector~\cite{PhysRevLett.121.086803}. Thus, the scaling $s_y$ grows exponentially with lattice index $y$ for directional hopping asymmetry, leading to exponential decay of $\psi_x(y)$ toward one boundary, which explains the NHSE observed numerically, even in the $\mathcal{PT}$-symmetric phase~\cite{PhysRevLett.116.133903,PhysRevLett.121.086803}.

\section{Effects of disorder on the NHSE SSH model}\label{sec5}

\subsection{Characterization measures of the NHSE}

To quantify the spatial confinement of the eigenstates, we use the inverse participation ratio (IPR), $R_{\rm IP}=\sum_{y=1}^{L}|\psi_x(y)|^4/(\langle\psi_x|\psi_x\rangle)^2$. For an extended state, $R_{\rm IP}\sim1/L$, and it vanishes in the thermodynamic limit, whereas $R_{\rm IP}\approx1$ indicates strong localization. Since the conventional IPR contains no information about the direction of localization, we further introduce the signed directional IPR (dIPR)~\cite{PhysRevB.105.245407,Li:24},
\begin{equation}\label{eq15}
R_{\rm dIP}(\psi_x)=\mathcal{P}(\psi_x)R_{\rm IP}(\psi_x).
\end{equation}
Here, $\mathcal{P}(\psi_x)$ is a polarization factor that identifies the spatial bias of the eigenstate,
\begin{equation}\label{eq16}
\mathcal{P}(\psi_x)=\mathrm{sgn}\left[\sum_{y=1}^{L}\left(y-\frac{L}{2}-\epsilon\right)|\psi_x(y)|\right],
\end{equation}
where $\epsilon$ is a small symmetry-breaking constant introduced to avoid ambiguity for nearly symmetric states. Accordingly, $\mathcal{P}=+1$ ($-1$) corresponds to localization biased toward the right (left) boundary. To characterize the overall directional localization of the spectrum, the directional mean IPR (dMIPR) is used by averaging the dIPR over all eigenstates, i.e.,
\begin{equation}\label{eq17}
R_{\rm dMIP}=\frac{1}{L}\sum_{x=1}^{L}R_{\rm dIP}(\psi_x).
\end{equation}

The topological robustness of the system in the presence of disorder is characterized by a real-space winding number. This quantity is constructed from the biorthogonal eigenbasis of the NH Hamiltonian $H_{\rm OBC}$ in Eq.~(\ref{eq12}). The right and left eigenvectors, denoted by $\{|\psi_x^R\rangle\}$ and $\{\langle\psi_x^L|\}$, are chosen to satisfy the biorthonormal condition $\langle\psi_{x'}^L|\psi_x^R\rangle=\delta_{x'x}$, which ensures the proper algebraic structure of the projection operators.

The chiral symmetry of the effective NH SSH chain is characterized by the operator
\begin{equation}\label{eqchiral}
S_{\rm ch}=\sum_{n=1}^{N}\left(\sigma_{n,a}^{+}|G_{2N}^{\rm eff}\rangle\langle G_{2N}^{\rm eff}|\sigma_{n,a}^{-}-\sigma_{n,b}^{+}|G_{2N}^{\rm eff}\rangle\langle G_{2N}^{\rm eff}|\sigma_{n,b}^{-}\right),
\end{equation}
which assigns opposite signs to the two sublattices $a$ and $b$ in each unit cell. Here, $|G_{2N}^{\rm eff}\rangle$ denotes the collective ground state of the effective NH SSH chain with $N$ unit cells, corresponding to $2N$ effective atomic sites. This operator satisfies the chiral-symmetry condition $\{S_{\rm ch},H_{\rm OBC}\}=0$. The flattened Hamiltonian is then constructed as $Q=\mathbb{I}-2P$, where $P=\sum_{x\in\rm occ}|\psi_x^R\rangle\langle\psi_x^L|$ is the biorthogonal projector onto the occupied subspace. The real-space winding number $\nu_o$ is obtained from the trace of the chiral displacement operator~\cite{PhysRevLett.113.046802,PhysRevLett.123.246801},
\begin{equation}\label{eq18}
\nu_o=\frac{1}{2N'}\mathrm{Tr}'\big(S_{\rm ch}Q[Q,X]\big),
\end{equation}
where $X$ is the coordinate operator and $[Q,X]$ encodes the spatial topological correlations between unit cells. The restricted trace $\mathrm{Tr}'$ is evaluated over a bulk window of length $N'=L-2l'$, excluding $l'$ sites near each boundary to suppress finite-size edge effects.

In realistic Rydberg-atom experiments, structural and control imperfections are unavoidable, including laser phase disorder, Rabi-frequency fluctuations, and position disorder. To characterize the response over multiple experimental realizations, we calculate the ensemble-averaged winding number over $N_s$ independent disorder configurations~\cite{PhysRevLett.123.246801,zhang2020non}:
\begin{equation}\label{eq19}
\nu=\frac{1}{N_s}\sum_{s=1}^{N_s}\nu_o^{(s)}.
\end{equation}
A value of $1-\nu\approx0$ indicates that the topological phase remains robust against the applied perturbations. This real-space invariant remains well-defined for finite systems under OBCs and provides a suitable diagnostic of topology in the presence of disorder~\cite{zhang2020non}. Unless otherwise specified, the parameters used in the following disorder analyses are the same as those in Figs.~\ref{fig1}--\ref{fig4}.

\subsection{Laser phase disorder}

\begin{figure}
\includegraphics[width=1\linewidth]{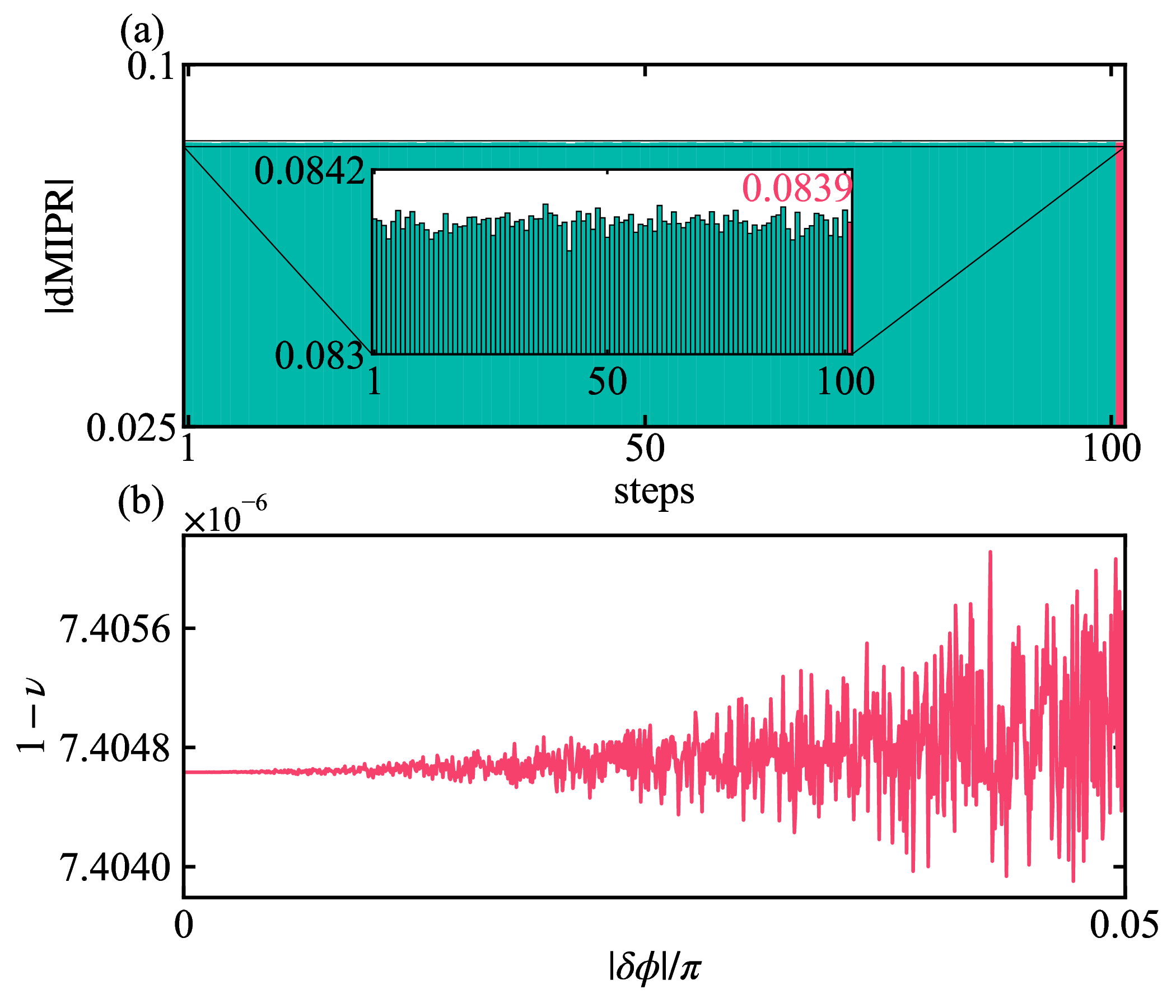}
\caption{(a) Robustness of the NHSE under phase disorder. Random-phase perturbations $\delta\phi$ sampled from a uniform distribution are introduced to the link phases $\phi^{ca}_{\rm III}=\pm\pi/2$ in a chain with 20 unit cells, and $|R_{\rm dMIP}|$ is evaluated according to Eq.~(\ref{eq20}) over $N_s=100$ independent disorder realizations. The green bars show the values obtained for individual realizations, while the red bar denotes the ensemble average. (b) Deviation of the winding number $1-\nu$ as a function of the disorder strength.}\label{fig5}
\end{figure}

Although the synthetic magnetic flux is, in principle, fixed by the lattice geometry and the prescribed laser-phase configuration, unavoidable experimental imperfections introduce small random perturbations to the enclosed flux. Physically, such phase disorder originates from local phase variations experienced by individual atoms due to residual thermal motion and relative phase drifts among the multiple dressing beams, which can be bounded within $\pi/20$ in current experiments~\cite{PhysRevLett.121.123603,PhysRevA.97.053803,PhysRevLett.124.033603}. To model this experimentally relevant effect, we introduce independent random-phase offsets $\delta\phi$ into these channels, leading to the effective Hamiltonian
\begin{eqnarray}\label{eq20}
H_{\delta\phi}&=&\sum_{n=1}^{L/2}\left(J_{L,{\rm ph}}\sigma^+_{2n-1}\sigma^-_{2n}+J_{R,{\rm ph}}\sigma^+_{2n}\sigma^-_{2n-1}\right)\nonumber\\
&&+\sum_{n=1}^{L/2-1}\left(G_{L,{\rm ph}}\sigma^+_{2n}\sigma^-_{2n+1}+G_{R,{\rm ph}}\sigma^+_{2n+1}\sigma^-_{2n}\right),
\end{eqnarray}
where
\begin{equation*}
J^{jk}_{\rm ph}=\left|\frac{\Omega_{\Theta^{[jk]}}^{j}\Omega_{\Theta^{[jk]}}^{k*}V^{jk}_{n,m}}{4\Delta_{\Theta^{[jk]}}\left(\Delta_{\Theta^{[jk]}}+V^{jk}_{n,m}\right)}\right|e^{i(\phi^{jk}_{\Theta^{[jk]}}+\delta\phi)},
\end{equation*}
with
\begin{eqnarray*}
J_{L,{\rm ph}}&=&J^{ab}+i\frac{4J^{bc}J^{ca}_{\rm ph}}{\Gamma},\qquad
J_{R,{\rm ph}}=J^{ab}-i\frac{4J^{bc}J^{ca}_{\rm ph}}{\Gamma},\\
G_{L,{\rm ph}}&=&J^{\rm inter}-i\frac{4\left(J^{bc}h_{1,{\rm ph}}+J^{ca}_{\rm ph}h_2\right)}{\Gamma},\\
G_{R,{\rm ph}}&=&J^{\rm inter}+i\frac{4\left(J^{bc}h_{1,{\rm ph}}+J^{ca}_{\rm ph}h_2\right)}{\Gamma}.
\end{eqnarray*}
Since the relevant gauge-invariant quantity is the total phase accumulated around a closed loop, phase fluctuations originating from different laser beams can be redistributed among the hopping links by a gauge transformation. Therefore, without loss of generality, we assign the effective flux disorder to the hopping channels associated with laser III, i.e., the $c\leftrightarrow a$ links. In Fig.~\ref{fig5}(a), phase disorder is implemented by assigning an independent random phase to each unit cell. The phase fluctuations are sampled from a uniform distribution, $\delta\phi\sim\mathcal{U}(-\eta,\eta)$, where $\eta$ is chosen to be 10\% of the Peierls phase $\pi/2$. Since the left- and right-boundary skin localizations are symmetric up to an overall sign, we use the absolute value $|R_{\rm dMIP}|$ to quantify boundary localization in a unified manner. Its value remains well above the delocalization limit, $1/L=0.025$, confirming that the localization persists under experimentally relevant phase noise. In Fig.~\ref{fig5}(b), we examine the topological invariant as a function of the disorder strength $|\delta\phi|$. As $|\delta\phi|$ increases, the deviation $1-\nu$ gradually departs from zero. Nevertheless, the real-space winding number $\nu$ remains robustly quantized, confirming the resilience of the topological phase against phase disorder. It is worth noting that the microscopic deviation $1-\nu\sim10^{-6}$ observed in the pristine system is purely a finite-size artifact, which vanishes in the thermodynamic limit and restores a perfectly quantized invariant.

\begin{figure}
\includegraphics[width=1\linewidth]{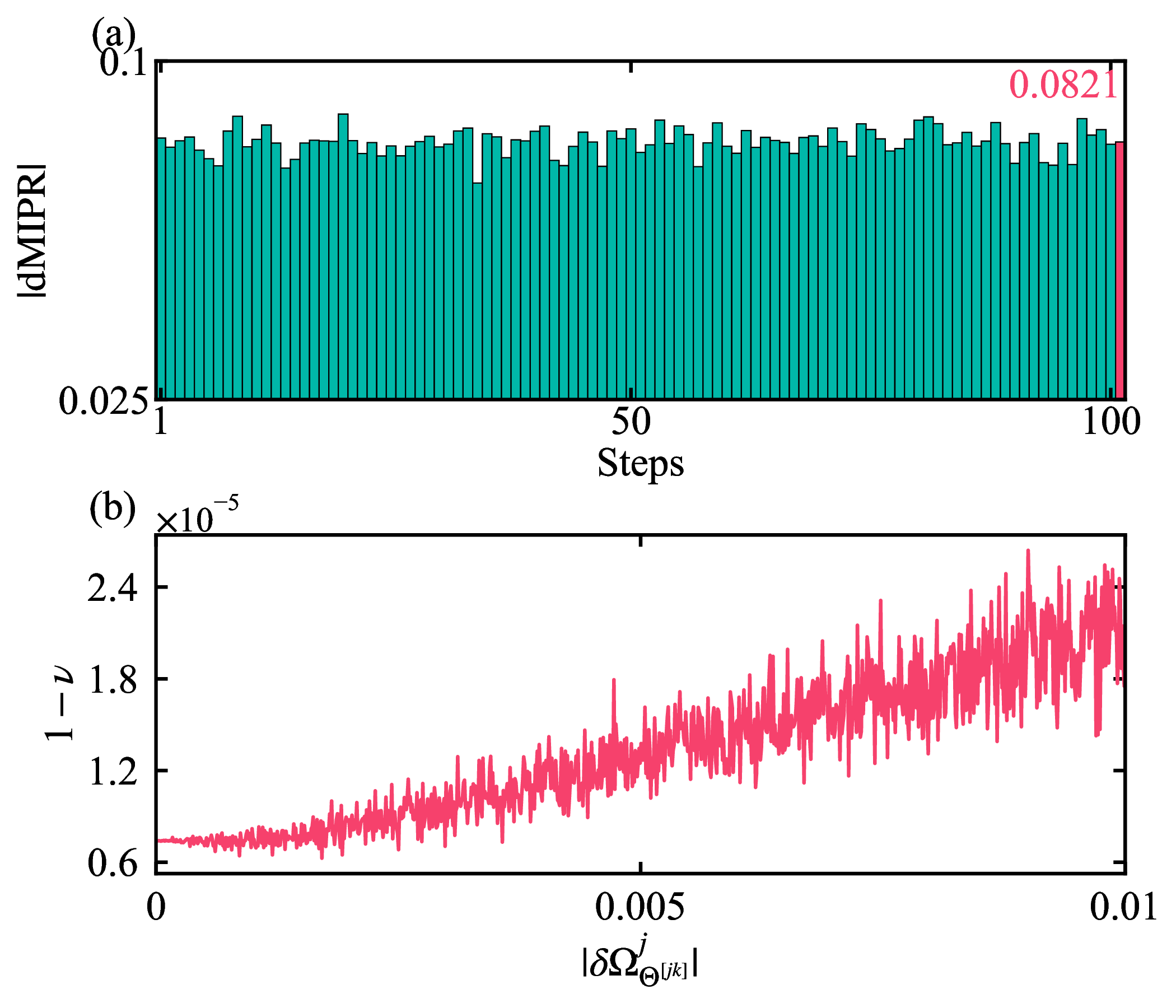}
\caption{(a) $|R_{\rm dMIP}|$ calculated from the effective Hamiltonian in Eq.~(\ref{eq21}) for $N_s=100$ independent realizations of Rabi-amplitude disorder. The green bars denote individual disorder realizations, while the red bar indicates the ensemble-averaged value. (b) Deviation of the winding number $1-\nu$ as a function of the Rabi-amplitude disorder strength.}\label{fig6}
\end{figure}

\begin{figure}
\includegraphics[width=1\linewidth]{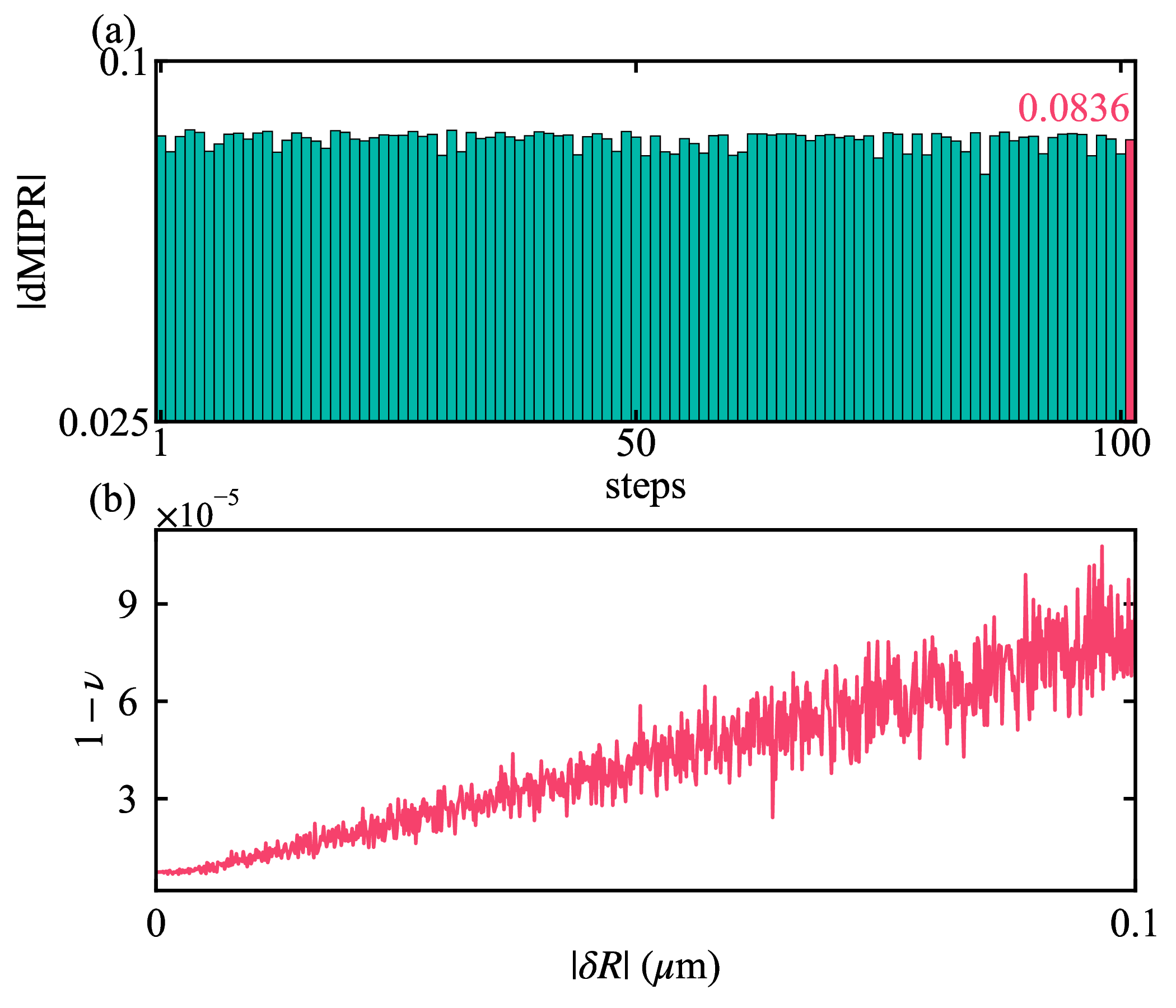}
\caption{(a) $|R_{\rm dMIP}|$ calculated from the effective Hamiltonian in Eq.~(\ref{eq22}) for $N_s=100$ independent realizations of positional disorder. The green bars denote individual disorder realizations, while the red bar indicates the ensemble-averaged value. (b) Deviation of the winding number $1-\nu$ as a function of the position disorder strength.}\label{fig7}
\end{figure}

\begin{figure*}
\includegraphics[width=1\linewidth]{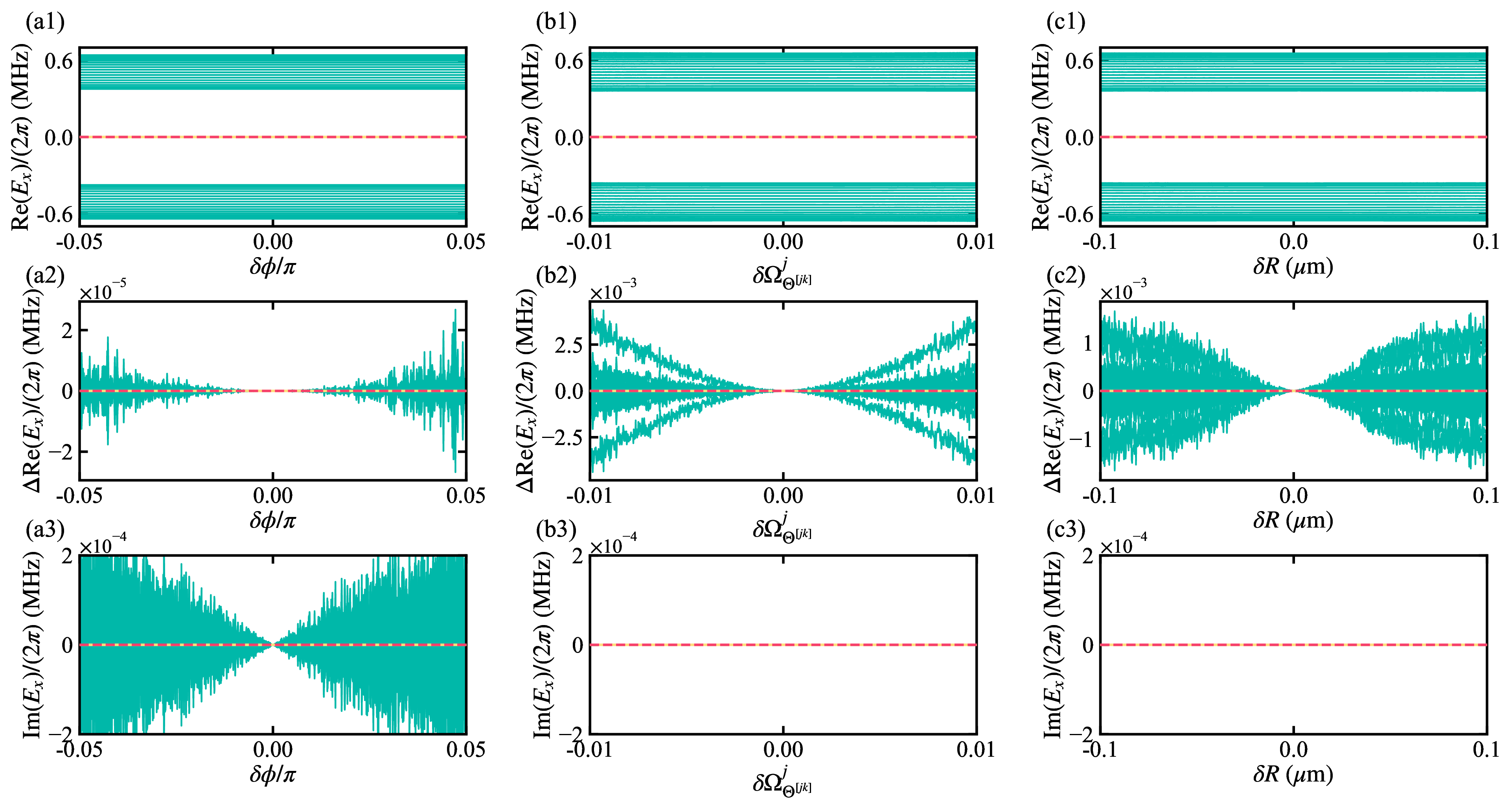}
\caption{Dependence of the complex eigenenergies on the laser phase fluctuation $\delta\phi/\pi$ [Eq.~(\ref{eq20})], the Rabi-amplitude fluctuation $\delta\Omega_{\Theta^{[jk]}}^{j}$ [Eq.~(\ref{eq21})], and the position fluctuation $\delta R~(\mu\mathrm{m})$ [Eq.~(\ref{eq22})]. (a1)--(c1) show the real parts of the energy spectra $\mathrm{Re}(E_x)/(2\pi)$~MHz, (a2)--(c2) display the disorder-induced shifts for all eigenstates, and (a3)--(c3) plot the imaginary parts $\mathrm{Im}(E_x)/(2\pi)$~MHz. The yellow and red lines represent the in-gap edge-state spectra, while the green lines denote the bulk-state spectra.}\label{fig8}
\end{figure*}

\subsection{Rabi-amplitude disorder}

After analyzing phase noise, we next consider fluctuations in the driving amplitudes. In realistic optical setups, residual beam inhomogeneity, slow power drifts, and intrinsic laser-intensity noise lead to stochastic variations of the Rabi frequencies, which can be modeled as Rabi-amplitude noise~\cite{PRXQuantum.4.020336,PhysRevA.111.022420,PRXQuantum.6.010331,7zjs-73qm}. We introduce this imperfection through $\Omega_{\Theta^{[jk]}}^{j\prime}=\Omega_{\Theta^{[jk]}}^{j}(1+\delta\Omega_{\Theta^{[jk]}}^{j})$, where $\delta\Omega_{\Theta^{[jk]}}^{j}$ is independently sampled from a bounded uniform distribution, $\delta\Omega_{\Theta^{[jk]}}^{j}\sim\mathcal{U}(-\eta',\eta')$. We set $\eta'=0.01$, corresponding to a maximum 1\% Rabi-amplitude deviation, and obtain the noisy effective Hamiltonian as
\begin{eqnarray}\label{eq21}
H_{\delta\Omega}&=&\sum_{n=1}^{L/2}\left(J_{L,{\rm ra}}\sigma^+_{2n-1}\sigma^-_{2n}+J_{R,{\rm ra}}\sigma^+_{2n}\sigma^-_{2n-1}\right)\nonumber\\
&&+\sum_{n=1}^{L/2-1}\left(G_{L,{\rm ra}}\sigma^+_{2n}\sigma^-_{2n+1}+G_{R,{\rm ra}}\sigma^+_{2n+1}\sigma^-_{2n}\right),
\end{eqnarray}
where
\begin{equation*}
J^{jk}_{\rm ra}=\Bigg|\frac{\Omega_{\Theta^{[jk]}}^{j'}\Omega^{k*'}_{\Theta^{[jk]}}V_{n,m}^{jk}}{4\Delta_{\Theta^{[jk]}}\big(\Delta_{\Theta^{[jk]}}+V_{n,m}^{jk}\big)}\Bigg|e^{i\phi^{jk}_{\Theta^{[jk]}}},
\end{equation*}
with
\begin{eqnarray*}
J_{L,{\rm ra}}&=&J^{ab}_{\rm ra}-\frac{4J^{bc}_{\rm ra}J^{ca}_{\rm ra}}{\Gamma},\qquad
J_{R,{\rm ra}}=J^{ab}_{\rm ra}+\frac{4J^{bc}_{\rm ra}J^{ca}_{\rm ra}}{\Gamma},\\
G_{L,{\rm ra}}&=&J^{\rm inter}_{\rm ra}+\frac{4\left(J^{bc}_{\rm ra}h_{1,{\rm ra}}+J^{ca}_{\rm ra}h_{2,{\rm ra}}\right)}{\Gamma},\\
G_{R,{\rm ra}}&=&J^{\rm inter}_{\rm ra}-\frac{4\left(J^{bc}_{\rm ra}h_{1,{\rm ra}}+J^{ca}_{\rm ra}h_{2,{\rm ra}}\right)}{\Gamma}.
\end{eqnarray*}
We further examine the stability of the NH topological response against Rabi-amplitude noise. Figure~\ref{fig6}(a) shows that the NHSE-induced boundary localization is preserved under realistic laser-intensity fluctuations. Meanwhile, Fig.~\ref{fig6}(b) confirms that the deviation $1-\nu$ remains close to zero for noise strengths up to 1\%, demonstrating that the topological invariant stays effectively quantized. These results show that the proposed SSH model retains both topological quantization and chiral nonreciprocal transport under experimentally relevant Rabi-amplitude disorder.

\subsection{Position disorder}

In realistic experiments, atomic positions are not perfectly fixed. Owing to zero-point motion and thermal fluctuations, an atom confined in an optical tweezer is more accurately described as a wave packet with a finite spatial extent. Depending on the trap depth and atomic temperature, typical position deviations range from $0.01$ to $0.1~\mu\mathrm{m}$~\cite{PhysRevLett.118.063606,PhysRevLett.132.223201,PhysRevX.15.011035,PhysRevResearch.7.L022035}. Since the vdW interaction depends sensitively on the interatomic distance, $V_{n,m}^{jk}\propto1/(R_{n,m}^{jk})^6$, even submicrometer displacements can produce appreciable variations in the effective coupling strengths. To incorporate this effect, we replace the ideal interatomic distance $R$ with $R+\delta R$, where $\delta R$ denotes a stochastic displacement. The resulting effective Hamiltonian becomes
\begin{eqnarray}\label{eq22}
H_{\delta R}&=&\sum_{n=1}^{L/2}\left(J_{L,{\rm sp}}\sigma^+_{2n-1}\sigma^-_{2n}+J_{R,{\rm sp}}\sigma^+_{2n}\sigma^-_{2n-1}\right)\nonumber\\
&&+\sum_{n=1}^{L/2-1}\left(G_{L,{\rm sp}}\sigma^+_{2n}\sigma^-_{2n+1}+G_{R,{\rm sp}}\sigma^+_{2n+1}\sigma^-_{2n}\right),
\end{eqnarray}
where
\begin{equation*}
J^{jk}_{\rm sp}=\Bigg|\frac{\Omega_{\Theta^{[jk]}}^{j}\Omega^{k*}_{\Theta^{[jk]}}V_{n,m}^{jk'}}{4\Delta_{\Theta^{[jk]}}\big(\Delta_{\Theta^{[jk]}}+V_{n,m}^{jk'}\big)}\Bigg|e^{i\phi^{jk}_{\Theta^{[jk]}}},
\end{equation*}
with $V_{n,m}^{jk'}=-C_6/(R+\delta R)^6$ in
\begin{eqnarray*}
J_{L,{\rm sp}}&=&J^{ab}_{\rm sp}-\frac{4J^{bc}_{\rm sp}J^{ca}_{\rm sp}}{\Gamma},\qquad
J_{R,{\rm sp}}=J^{ab}_{\rm sp}+\frac{4J^{bc}_{\rm sp}J^{ca}_{\rm sp}}{\Gamma},\\
G_{L,{\rm sp}}&=&J^{\rm inter}_{\rm sp}+\frac{4\left(J^{bc}_{\rm sp}h_{1,{\rm sp}}+J^{ca}_{\rm sp}h_{2,{\rm sp}}\right)}{\Gamma},\\
G_{R,{\rm sp}}&=&J^{\rm inter}_{\rm sp}-\frac{4\left(J^{bc}_{\rm sp}h_{1,{\rm sp}}+J^{ca}_{\rm sp}h_{2,{\rm sp}}\right)}{\Gamma}.
\end{eqnarray*}
In Fig.~\ref{fig7}(a), independent random perturbations are applied to all interatomic distances, with each displacement sampled from a uniform distribution $\delta R\in[-0.1,0.1]~\mu\mathrm{m}$. The ensemble-averaged localization measure remains sizable, with $|R_{\rm dMIP}|\approx0.0836$, indicating that moderate position fluctuations do not suppress the NHSE-induced boundary localization or the associated nonreciprocal transport asymmetry. Figure~\ref{fig7}(b) further shows that the deviation $1-\nu$ stays close to zero throughout the entire disorder range, confirming that the topological invariant remains effectively quantized. These results demonstrate that the proposed SSH model retains its NH topological response under realistic geometric imperfections.

\subsection{Response of the spectrum to disorder}

We finally examine the response of the eigenspectrum to phase disorder $\delta\phi$, Rabi-amplitude disorder $\delta\Omega_{\Theta^{[jk]}}^{j}$, and position disorder $\delta R$. Figures~\ref{fig8}(a1)--\ref{fig8}(c1) display the real parts of all eigenvalues, highlighting the two in-gap edge states. Figures~\ref{fig8}(a2)--\ref{fig8}(c2) show the disorder-induced shifts, $\Delta\mathrm{Re}(E_x)=\mathrm{Re}[E_x(\delta)]-\mathrm{Re}[E_x(0)]$, for all eigenstates. These plots reveal that the bulk modes exhibit small fluctuations, while the edge states remain tightly pinned near zero. Figures~\ref{fig8}(a3)--\ref{fig8}(c3) present the imaginary parts of the eigenvalues, demonstrating that only phase disorder generates finite $\mathrm{Im}(E_x)$, whereas Rabi-amplitude and position disorder preserve a purely real spectrum within the considered ranges $\delta\Omega_{\Theta^{[jk]}}^{j}\leq0.01$ and $\delta R\leq0.1~\mu\mathrm{m}$. These results confirm that the topological edge states are stable against experimentally relevant phase, amplitude, and position imperfections.

\section{Generalization to the periodic-boundary NH SSH model}\label{sec6}

We now extend the system to PBCs. In contrast to the open-chain geometry, PBCs remove physical boundaries and therefore eliminate boundary-induced localization. The topological characterization must then rely on a bulk quantity defined directly in real space.

\begin{figure}
\includegraphics[width=0.9\linewidth]{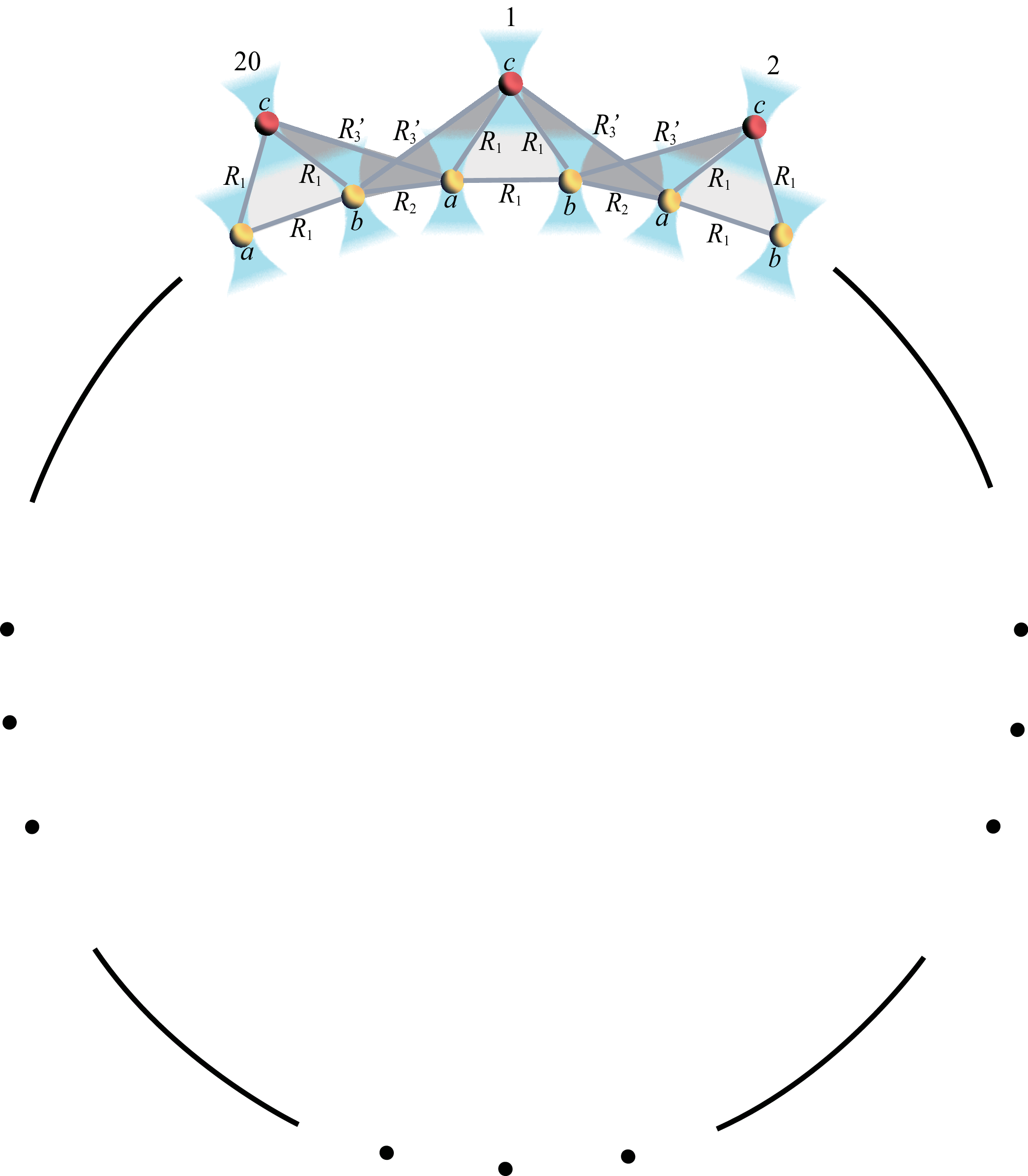}
\caption{Schematic of the PBC geometry formed by arranging $N=20$ three-atom unit cells into a ring.}\label{fig9}
\end{figure}

As shown in Fig.~\ref{fig9}, the PBC geometry is realized by arranging 20 unit cells into a ring. Using the same unit-cell labeling as in the OBC chain, PBCs are implemented by adding a nonreciprocal intercell hopping channel between atom $a$ of the first labeled unit cell and atom $b$ of the last labeled unit cell. This additional link closes the chain and removes the physical boundaries. The effective intra- and intercell coupling strengths are kept the same as in the open-chain case, while a larger separation $R'_3\approx8.61~\mu\mathrm{m}$ is used to suppress unwanted cross-couplings for $R>R'_3$. Under these conditions, the effective Hamiltonian with PBCs reads
\begin{eqnarray}\label{eq23}
H_{\rm PBC}&=&\sum_{n=1}^{L/2}\left(J_L\sigma^+_{2n-1}\sigma^-_{2n}+J_R\sigma^+_{2n}\sigma^-_{2n-1}\right)\nonumber\\
&&+\sum_{n=1}^{L/2-1}\left(G_L\sigma^+_{2n}\sigma^-_{2n+1}+G_R\sigma^+_{2n+1}\sigma^-_{2n}\right)\nonumber\\
&&+G_L\sigma^+_L\sigma^-_1+G_R\sigma^+_1\sigma^-_L.
\end{eqnarray}

To characterize the bulk topology of the disordered NH system under PBCs without invoking momentum space, we construct a real-space topological invariant based on the Bott-index idea~\cite{PhysRevLett.80.1800,PhysRevB.103.224208} and the biorthogonal framework of NH systems. We first solve the right and left eigenstates, $|\psi_z^R\rangle$ and $\langle\psi_z^L|$, satisfying the biorthogonal normalization condition $\langle\psi_{z'}^L|\psi_z^R\rangle=\delta_{z'z}$. The flattened spectral operator is then defined as
\begin{equation}
Q'=\mathbb{I}-2\sum_{z=1}^{N}|\psi_z^R\rangle\langle\psi_z^L|.
\end{equation}
Because of chiral symmetry, $Q'$ can be written in the off-diagonal form
\begin{equation}
Q'=\begin{pmatrix}0&q\\q^{-1}&0\end{pmatrix},
\end{equation}
where the block $q$ encodes the chiral topological information. For PBCs, the usual linear position operator is not suitable because it is discontinuous at the boundary. We therefore introduce the periodic unitary position operator
\begin{equation}
U_X=\exp\left(i\frac{2\pi}{N}\hat{X}\right),
\end{equation}
with $\hat{X}=\mathrm{diag}(0,1,2,\ldots,N-1)$. The corresponding real-space NH winding number is then defined as
\begin{equation}
\nu_p=\frac{1}{2\pi i}\operatorname{Tr}\ln\left(q^{-1}U_XqU_X^\dagger\right).
\end{equation}
For disordered systems, we average this quantity over $N_s$ independent disorder realizations,
\begin{equation}
\nu'=\frac{1}{N_s}\sum_{s=1}^{N_s}\nu_p^{(s)}.
\end{equation}
This construction provides a real-space bulk diagnostic for the PBC system and remains applicable when translational symmetry is broken by phase disorder, Rabi-amplitude disorder, or positional disorder. We have also checked numerically that, within the experimentally relevant parameter ranges considered above, the real-space NH topological invariant remains quantized under all three types of disorder. This consistency check confirms that the bulk topology of the proposed NH SSH model is not altered by these realistic perturbations.

\section{Conclusion}\label{sec7}

In summary, we proposed a real-space Rydberg-atom-array platform for realizing an NH SSH model with controllable nonreciprocal couplings. The system consists of three-atom unit cells, where auxiliary atoms are introduced to mediate effective hopping channels. By engineering the dissipative dynamics and adiabatically eliminating the auxiliary degrees of freedom, we obtained an effective two-sublattice SSH Hamiltonian with asymmetric intra- and intercell couplings. This construction provides a direct and experimentally accessible route to NH topology in Rydberg arrays.

We characterized the resulting NH topological properties under OBCs. The system exhibits a pronounced NHSE and in-gap edge states, whose eigenenergies remain pinned near zero under realistic perturbations. The skin-effect order parameter and the real-space topological invariant further confirm the stability of the nonreciprocal topological response. We examined several experimentally relevant imperfections, including laser phase noise, Rabi-amplitude fluctuations, and position disorder. The results show that the boundary localization, topological quantization, and edge-state spectrum remain robust within the parameter ranges accessible to current Rydberg-atom experiments. Building on this, we further extended the model to PBCs, enabling the investigation of bulk topological properties using a real-space NH winding number.

A key feature of our scheme is that the effective SSH model is realized directly in real space, rather than through synthetic dimensions or momentum-space constructions. This real-space setting allows us to discuss both open- and periodic-boundary geometries within the same physical platform. In particular, we employ a real-space NH winding number based on biorthogonal projectors to characterize the bulk topology. This approach is especially useful in the presence of disorder, where translational symmetry is broken and conventional momentum-space winding numbers are no longer directly applicable.

Given the strong and tunable interactions of Rydberg atoms, together with the controllability of coherent driving and engineered dissipation, our proposal provides a flexible platform for programmable NH topological dynamics. It may also be extended to interacting topological chains, higher-dimensional NH lattices, and dissipative many-body systems, offering opportunities to explore the interplay among topology, nonreciprocity, disorder, and interactions in controllable atomic arrays.

\begin{acknowledgments}
The authors thank the anonymous referees for their careful reading and constructive suggestions, which helped improve the clarity and presentation of the manuscript. This work was supported by the National Natural Science Foundation (Grants No. 12174048 and No. 12564047). W.L. acknowledges support from the EPSRC through Grant No. EP/W015641/1. F.Y. acknowledges support from the Firuza Foundation Fellowship and helpful discussions with B. Yan and X. Wu.
\end{acknowledgments}

\section*{Data Availability}
The data that support the findings of this article are openly available~\cite{bai_2026_21438786}, embargo periods may apply.

\bibliography{manuscript.bbl}

\end{document}